# Observation of the orbital quantum dynamics in the spin-1/2 hexagonal antiferromagnet $Ba_3CuSb_2O_9$

Yibo Han[1,2], Masayuki Hagiwara[1*], Takehito Nakano[3], Yasuo Nozue[3], Kenta Kimura[4], Mario Halim[4], Satoru Nakatsuji[4,5]

[1]Center for Advanced High Magnetic Field Science, Graduate School of Science, Osaka University, 1-1 Machikaneyama, Toyonaka, Osaka 560-0043, Japan

[2]Wuhan National High Magnetic Field Center, Huazhong University of Science and Technology, Wuhan 430074, China

[3]Graduate School of Science, Osaka University, 1-1 Machikaneyama, Toyonaka, Osaka 560-0043, Japan

[4]Institute for Solid State Physics, University of Tokyo, 5-1-5 Kashiwanoha, Kashiwa, Chiba 277-8581, Japan

[5]PRESTO, Japan Science and Technology Agency (JST), 4-1-8 Honcho Kawaguchi, Saitama 332-0012, Japan

**Abstract:** We explore orbital dynamics in the spin liquid candidate $Ba_3CuSb_2O_9$ using multi-frequency electron spin resonance. We prepared two high quality single crystals. The crystal with a slight copper deficiency shows a structural phase transition at around 200 K due to the cooperative Jahn-Teller effect, accompanied with orbital ordering. In contrast, the crystal with almost perfect stoichiometry shows no orbital ordering down to the lowest temperature of 1.5 K. Dramatic change in the *g*-factor anisotropy as a function of frequency and temperature demonstrates orbital quantum fluctuations at a nearly constant time scale of ~ 100 ps below 20 K, evidencing the emergence of an orbital liquid state in this quantum spin liquid compound.

PACS codes: 71.70.Ej, 75.10.Kt, 75.25.Dk, 76.30.-v

Extensive work on various quantum spin liquid candidates has revealed that dynamic spin state may emerge close to zero Kelvin in Mott insulators with frustrated interactions between the spins [1]. Question arises whether an analogous state is possible if additional degrees of freedom are present. In the case of orbital degeneracy, theoretical investigations revealed various types of quantum liquids [1-6]. However, the experimental realization of such states is still a challenge [7-10]. For a strongly correlated system composed of orbital-degenerated 3$d$ transition ions (such as $Cu^{2+}$, $Mn^{3+}$, $Cr^{2+}$) in an octahedral cubic crystalline field, it is commonly known that the orbital degrees of freedom is mostly quenched by a *Jahn-Teller* (JT) effect where the JT transition occurs accompanied by crystal symmetry lowering and orbital long-range ordering at a fairly high temperature, a few thousands Kelvin, which is at least one order of magnitude higher in energy than the magnetic *exchange* interaction [11].

An exception was found for the hexagonal (6H) perovskite quantum spin liquid candidate: $Ba_3CuSb_2O_9$ (BCSO) [12-13]. In this material, a weakly connected honeycomb-based network of JT active $CuO_6$ octahedra face-sharing with JT inactive $SbO_6$ rigid octahedra allows the energy scales of the JT and exchange interaction to become comparable in strength. Various measurements have revealed that the hexagonal BCSO exhibits neither magnetic long-range order nor cooperative JT transition [13-15]. The absence of orbital long-range order allows the studies on the formation of quantum spin and/or orbital liquid state in this material.

A key question is to identify the nature of the spin and orbital dynamics in BCSO, whether spatial or temporal orbital disorders exist at low temperatures. For orbital dynamics, two distinct scenarios are possible: (I) static JT distortions appear locally, forming a glass type of orbital freezing, and realizing a rare case of random-singlet state [16-17]; (II) the distortions are dynamical, with a finite tunneling rate between the orbital-vibronic states at the local minima of an effective potential describing the JT coupling in the octahedra –the so called dynamical JT effect [13-14, 18-19]. This may lead to a quantum orbital liquid state. If the latter is true, the characteristic frequency of the dynamic JT distortions must be higher than $10^{10}$ Hz according to our previous K-band electron spin resonance (ESR) and extended X-ray absorption fine structure (EXAFS) observations [13]. The specific heat, neutron scattering, muon spin and nuclear magnetic resonance measurements did not confirm the above scenarios, probably because the timescales in these measurements are very different from the timescale of the orbital fluctuations, and further because the imperfect hexagonal polycrystalline samples complicated the analysis [13, 16].

To this end, we succeeded in preparing two kinds of high-quality single crystals of $Ba_3Cu_{1-x}Sb_{2+x}O_9$ with distinct structural behaviors [14]. One named "orthorhombic sample" with slight copper deficiency ($x = 0.042(1)$) shows a structural transition at about 200 K between a high-temperature hexagonal ($P6_3/mmc$) and a low-temperature orthorhombic ($Cmcm$) phases, exhibiting a cooperative JT transition. The other, named "hexagonal sample", with almost perfect Sb/Cu stoichiometry ($x < 0.003$) remains hexagonal at all experimental temperatures, serving as the first copper oxide without the cooperative JT transition. Here we chose multi-frequency ESR measurement technique to investigate these samples, because its dynamic range could be extended to high frequency region to resonantly detect the spin and orbital magnetism simultaneously.

Let us first present the X-band ESR results at $\nu_{EM} \sim 9$ GHz, the lowest frequency available with our instrument. Distinct behaviors in the angular dependent ESR spectra are observed for the orthorhombic (Fig. 1(a)) and hexagonal (Fig. 1(b)) samples at 2.4 K. For further discussion, we derived the *g*-factor, ESR intensity, and linewidth from the fitting of the curves using field-derivative multiple or single Lorentzian functions [20].

The *g*-factors of the orthorhombic sample clearly show anisotropic behavior due to static JT distortions (Fig. 1(c)). When the magnetic field was rotated in the *ab*-plane, three main branches with the period of 180° were observed, each branch shifted by 60° from one another, indicating the formation of three kinds of CuO$_6$ domains with orthorhombic elongation along the respective three Cu-O bond directions. The characteristic *g*-factors along two long and four short Cu-O bonds are estimated as $g_\parallel$= 2.41(1) and $g_\perp$= 2.08(1) [20], which are standard values known for a strong static JT distortion in a CuO$_6$ octahedron [21]. On heating, a transition of the in-plane *g*-factor from anisotropic to isotropic was observed at ~ 200 K (Fig. 1(e)), confirming the cooperative JT transition.

For the hexagonal sample, however, even the lowest temperature (2.4 K) ESR curves in Fig. 1(b) retain a perfectly symmetric Lorentzian line shape [20]. It is more striking that the *g*-factor is almost isotropic (with a tiny six-fold symmetry) along the in-plane directions as shown in Figs. 1(d) and 1(f), revealing the absence of cooperative JT distortion. This is unprecedented for a cupric single crystal at such a low temperature. The in-plane *g*-factor anisotropy at $T > 20$ K (Fig. 1(f)) is totally smeared out similarly to the case of the orthorhombic sample at $T > 200$ K (Fig. 1(e)), which is most likely due to the thermally activated dynamic JT effect for the electrons coupled with the vibronic excited states [22-23].

Concerning spin dynamics, we have investigated the X-band ESR intensity and linewidth. The ESR intensities in Fig. 2(a) show similar behavior for both samples, which are proportional to the static susceptibilities [24]. At low temperatures Curie-Weiss behavior is shown for both samples, which can be attributed to about 14% "orphan spins" in the hexagonal sample, and 18% in the orthorhombic one. Subtracting the Curie-Weiss component, the remaining ESR intensity data can be fitted with a formula for the "dimer" model in which the singlet ground state and the triplet excited one are separated by the gapped energy ~ 70 K [23]. The behavior indicates most copper spins form singlet dimers below the exchange interaction scale $J/k_B$ ~50 K [16].

More information about the couplings of the spin, orbital and lattice could be obtained from the ESR linewidth (FWHM, Full Width at Half Maximum) as shown in Fig. 2(b). Both samples show a linewidth broadening with increasing temperature except for the low temperature region of the orthorhombic sample along the [100] direction. The linewidth is much broader in the hexagonal sample than in the orthorhombic one, and a plateau (temperature independent region) appears for the hexagonal sample, while not for the orthorhombic one. The anisotropic exchange interaction and the thermal effect may determine the ESR linewidth broadening: FWHM = $K_{AE}\exp[-C_1/(T+C_2)] + K_{TE}\exp(-\Delta_{ESR}/k_B T)$ [25]. By fitting the FWHM data along the *c*-axis for both samples with this equation and extracting the coefficients and the parameter values, we have found that the coefficient $K_{AE}$ for the hexagonal sample is twice as large as that of the orthorhombic one. In contrast, the coefficients $K_{TE}$ for both samples are similar, and the energy gaps are evaluated to be $\Delta_{ESR}/k_B$ = 444 K for the hexagonal sample and $\Delta_{ESR}/k_B$ = 568 K for the orthorhombic one, indicating a similar thermal fluctuation at high temperatures.

The above X-band ESR results suggest a low temperature orbital ordering in the orthorhombic sample and the orbital fluctuations in the hexagonal one [20]. However, more convictive evidence is necessary for corroborating that this is caused by temporal orbital fluctuating rather than spatial orbital disordering. High frequency-high field ESR spectroscopy which covers the frequencies of $10^{10} \sim 10^{12}$ Hz could be a key experiment to delineate the possible orbital dynamics. The dynamic orbital fluctuations should manifest themselves as a crossover from an isotropic to an anisotropic line shape with increasing frequency due to two kinds of mechanisms. (I) Snapshot effect: when the period of the electromagnetic wave is much longer than the time scale of orbital tunneling $\tau$, a motional narrowed and isotropic ESR signal could be observed. On the contrary, when $1/\nu_{EM} << \tau$, the magnetic moment become "slowly motional", and consequently, the ESR line shape would approach that of the static JT distortion [23]. (II) Quenching of orbital quantum fluctuations under high magnetic fields [22-23]: Due to spin-orbit coupling, the magnetic field and the distortions are coupled via the anisotropic $g$-factor. A magnetic field therefore acts as a local strain. If the potential energy of this strain, which is proportional to the cost $g_2\mu_B H$ of the anisotropy in the Zeeman energy, where $g_2 = g_\parallel - g_1$, and $g_1$ is the average of the $g$-values [20], overcomes the dynamic energy due to tunneling between the local minima, the orbitals will freeze and static JT distorted ESR spectra are expected.

Significantly, our high-frequency ESR results are found consistent with the above predictions. Figure 3 shows that the line shapes of the ESR absorption curves are clearly $\nu_{EM}$-dependent. In particular, the spectra become deformed from a symmetric Lorentzian curve when $\nu_{EM}$ exceeds a critical value: $\nu_C \sim 80$ GHz at 1.5 K (Figs. 3(a), (b)), and $\nu_C \sim 180$ GHz at 50 K (Figs. 3(c), (d)). Notably, the ESR line shape at $\nu_{EM} = 730$ GHz (resonance field $H_0 = 23.7$ T for the average $g$-value $g_1 = 2.2$) for the hexagonal sample is almost the same as the orthorhombic sample's case at 9.3 GHz, providing a high-frequency "snapshot" of the dynamic JT effect. Figure 2(a) shows that at 50 K, a temperature comparable to the spin gap, the ESR probes all the Cu sites. Therefore, the observed crossover confirms that the dynamic JT effect is a *bulk* effect.

We also note that our results clearly rule out the exchange splitting effect (ESE) [29], the other popular mechanism for the frequency dependent crossover in ESR. Within the ESE scenario, one would assume that the local copper sites are in an orbital glass state and the exchange energy between the orphan spins is $J'/k_B = \Delta g \mu_B H_0 \sim 0.5$ K ($\Delta g = g_\parallel - g_\perp = 0.32$ at the highest frequency, and $H_0 = 26.0$ kOe is the resonance field at $g_1 = 2.2$ and $\nu_C = 80$ GHz). However, two following facts dispel this possibility: Firstly, when plotted as a function of $\Delta H/H_0$, the single Lorentzian line width decreases with increasing frequency at $\nu_{EM} < \nu_C$, which is opposite to the tendency expected for the ESE scenario [29]. Secondly, $\nu_C$ shifts to higher frequencies on heating (see Fig. 3), which is also opposite to the tendency expected for the ESE: $\nu_C$ would decrease because the exchange interaction would become weaker as the lattice expands [13]. Instead, the broad line width at low frequencies can be naturally understood as the results of dynamic mixing of *g*-tensor, namely, the low frequency snapshot; and the observed increase of $\nu_C$ with temperature can be understood by the fact that the orbital tunneling speeds up due to thermal fluctuations.

Thus, the dramatic change may well come from the dynamic orbital fluctuations. Using single ion approximation [28], we estimate the characteristic frequency of the dynamic JT effect $\nu_{JT}$ by the relation $\nu_{JT} = \nu_C g_2/g_1 \approx 0.095*\nu_C$ [20]. As shown in Fig. 4(a), $\nu_{JT}$ systematically decreases on cooling, indicating the suppression of thermal orbital fluctuations. However, when the temperature gets lower than 20 K, $\nu_{JT}$ levels off at a constant value ~ 10 GHz, suggesting that the orbital quantum tunneling has the timescale of about 100 picoseconds. The temperature dependence of $\nu_{JT}$ defines the three essentially different regimes: (i) At the high-frequency-low-temperature limit, the *g*-factor becomes as anisotropic as the orthorhombic sample's case with the static distortion as shown in Fig. 4(b). This identifies the strength of dynamic JT distortion is comparable with that of the static JT effect, consistent with the previous EXAFS results [13]. (ii) At the low-frequency-high-temperature limit, the fully isotropic *g*-factor due to thermal orbital fluctuations is shown in Fig. 4(e). (iii) In the low-frequency-low-temperature "quantum tunneling" region, the six-fold anisotropy of the *g*-factor (Fig. 4(c)) was observed at $\nu_{EM} \leq 80$ GHz and $T \leq 20$ K.

This emergent six-fold g-factor anisotropy cannot be understood simply based on the theory for the single-ion dynamic JT effect [27-28]. Is it the fingerprint of a spin-orbital entangled state, where Cu spins form fluctuated singlet dimers which drive the dynamical Jahn Teller effect? Or is it just coming from the quantum tunneling between the orbital minima within a single ion? Various theoretical models have been developed for the BCSO:

the electron-vibronic effects are emphasized in Refs. [18-19], while in Ref. [30], the importance of the coupling between the Cu spins on the honeycomb lattice and the decorating sites is stressed. It remains to be seen if these theories are compatible with our experimental findings, including the explanation of the behavior of the orphan spins, especially since we observe the same quantum fluctuations for orbitals in the bulk and on the orphan sites.

To conclude, our multifrequency ESR measurements have identified two regimes for the dynamical Jahn-Teller effect, in addition to the cooperative Jahn-Teller effect, as a function of temperature and magnetic field. Most notably, the characteristic frequency of the orbital fluctuation below 20 K in the hexagonal BCSO remains finite and temperature independent. Together with the six-fold symmetry of the g-factor, this indicates disordered orbitals with strong quantum fluctuations, forming a quantum orbital liquid. This is a sharp contrast with the static JT effect and orbital ordering observed in the orthorhombic BCSO. Together with the previous X-ray and Raman examinations [14], we believe that this orbital liquid is realized not only at high temperatures ($T \geq 50$ K), but also at low temperatures ($T < 50$ K), although additional experiments such as infrared optical spectroscopy are required to further confirm this point. Finally, it is interesting to note that the frequency-dependent ESR curves in Fig. 3 are analogous to those of free spins "swimming" in a liquid solution in the bio-chemical "spin labels" [31], which offers a classical image of the ESR response from the orphan spins itinerating in the bulk orbital liquid.

We thank H. Sawa, S. Ishihara, Y. Wakabayashi, C. Broholm, N. Katayama, K. Kuga, and J. Nasu for discussions, K. Penc for critical reading of the manuscript and useful suggestions, and T. Kida and T. Fujita for the assistance in the pulsed high field ESR experiments. This research is partly supported by Grants-in-Aid for Scientific Research (No. 242440590 and 25707030), the Global COE Program (Core Research and Engineering of Advanced Materials and Interdisciplinary Education Center for Materials Science, Grant No.

G10), both from the MEXT, Japan, by PRESTO of JST, Japan, and by the National Science Foundation of China (No. 11104097). The use of the facilities of the Materials Design and Characterization Laboratory at the Institute for Solid State Physics, The University of Tokyo, is gratefully acknowledged. Some of these studies were done under a Foreign Visiting Researcher Program in the Center for Quantum Science and Technology under Extreme Conditions (KYOKUGEN).

**Figure caption:**

FIG. 1 (color online). X-band ESR results measured at 9.3 GHz. (a) ESR curves of the hexagonal sample (Hexa.) for external magnetic fields rotating in *ab* plane at $T = 2.4$ K, (b) ESR curves for the orthorhombic sample (Ortho.) with the same setting. (c)-(d) In-plane *g*-factor anisotropy obtained from the fits of the curves in (a)-(b), solid lines are fits to the data by the sinusoidal function, inset of (d) defines the rotating angles $\varphi$. (e)-(f), Temperature-dependent *g*-factors at $\varphi = 0°$ and 30°. Note that the *g*-factor scale is much different between (c) (e) and (d) (f).

FIG. 2 (color online). (a) ESR intensities for the hexagonal and the orthorhombic samples measured at 9.3 GHz. The "Dimer" components are extracted by subtracting the "Curie-Weiss" components (low temperature fit) from the total intensity. The obtained Weiss temperatures are nearly 0 K for both samples. The expression of dimer fittings is found in the SM5 [20]. (b) ESR linewidths for both samples along the *a*- and *c*-axes. Solid lines are the total fits, and broken lines are the anisotropic exchange fits.

FIG. 3 (color online). Multi-frequency ESR absorption spectra. (a) $\varphi = 0°$, $T = 1.5$ K, (b) $\varphi = 30°$, $T = 1.5$ K, (c) $\varphi = 0°$, $T = 50$ K, (d) $\varphi = 30°$, $T = 50$ K. The microwave frequency $\nu_{EM}$ varies from 9.3 to 730 GHz for the hexagonal sample. The spectra of the orthorhombic one at $\nu_{EM} = 9.3$ GHz are shown on top. $\Delta H/H_0 = (H-H_0)/H_0$, $H_0 = h\nu_{EM}/(g_1\mu_B)$, $g_1 = 2.2$. The sharp peaks at $\Delta H/H_0 \approx 0.1$ for $\nu_{EM} = 730$ GHz are the signals of DPPH. Open circles show the data, and broken and solid curves are the multi-peak Lorentzian fits and the total fits, respectively.

FIG. 4 (color online). (a) Temperature dependence of the Jahn-Teller frequency ($\nu_{JT}$). (b)

Two-fold symmetric $g$-factor at $\nu_{EM} = 380$ GHz, $T = 1.5$ K. (c) Six-fold symmetric $g$-factors at $\nu_{EM} = 80$ GHz, $T = 1.5$ K. (d) Tiny six-fold symmetric $g$-factors at $\nu_{EM} = 9.3$ GHz, obtained at $T = 2.4$ K and 10 K. (e) Isotropic $g$-factor at $\nu_{EM} = 9.3$ GHz, obtained at $T = 30$ K and 100 K. The experimental conditions (frequencies and temperatures) for (b)-(e) are shown in (a) as corresponding symbols.

FIG. 1

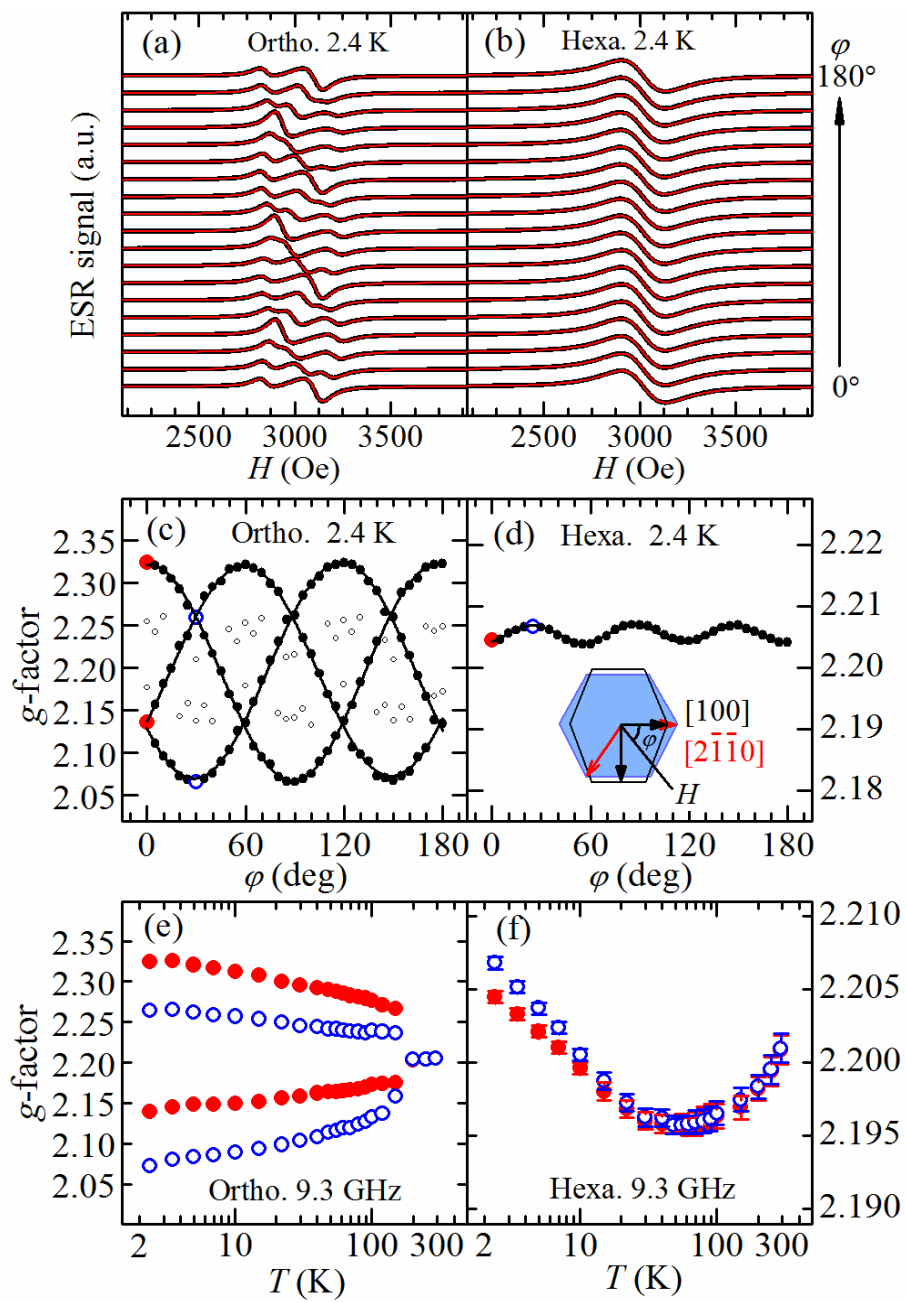

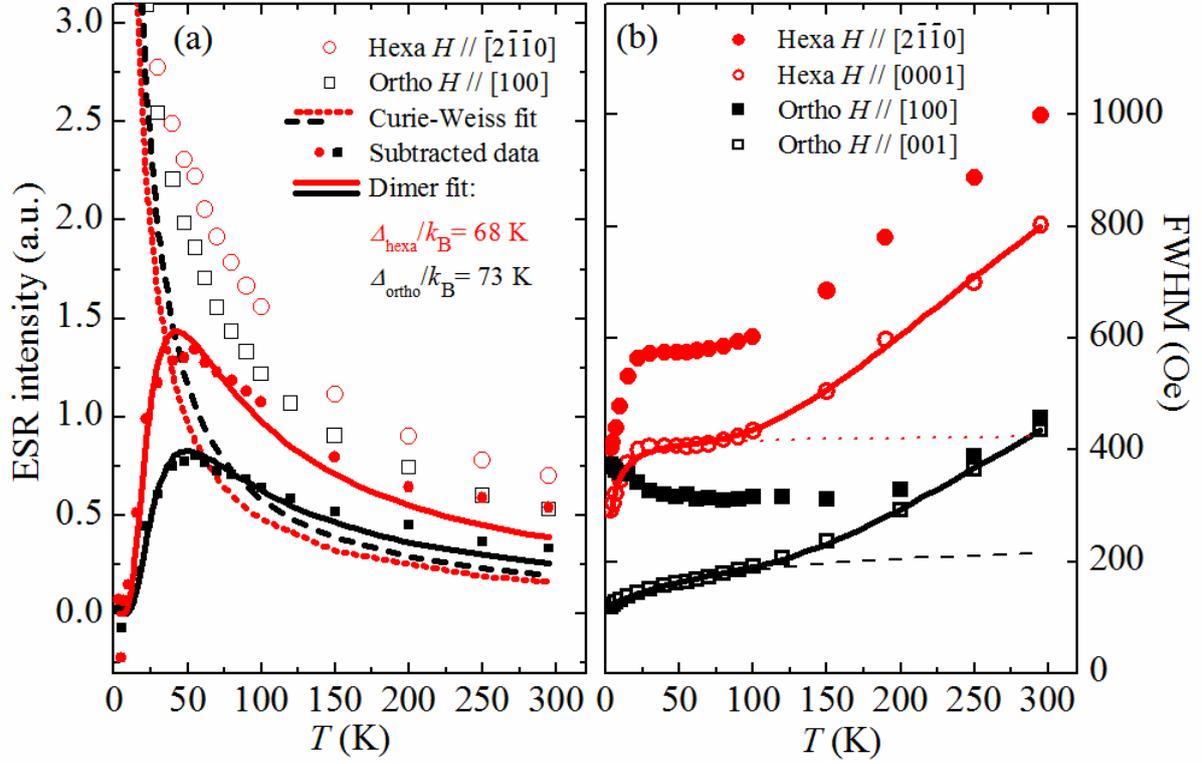

FIG. 3

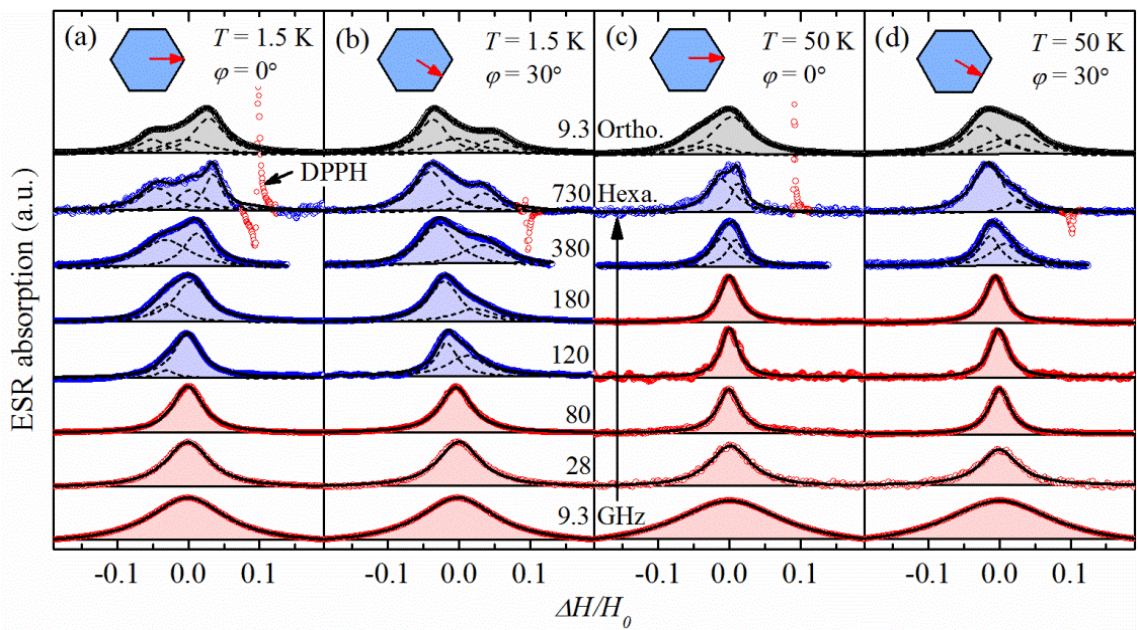

FIG. 4

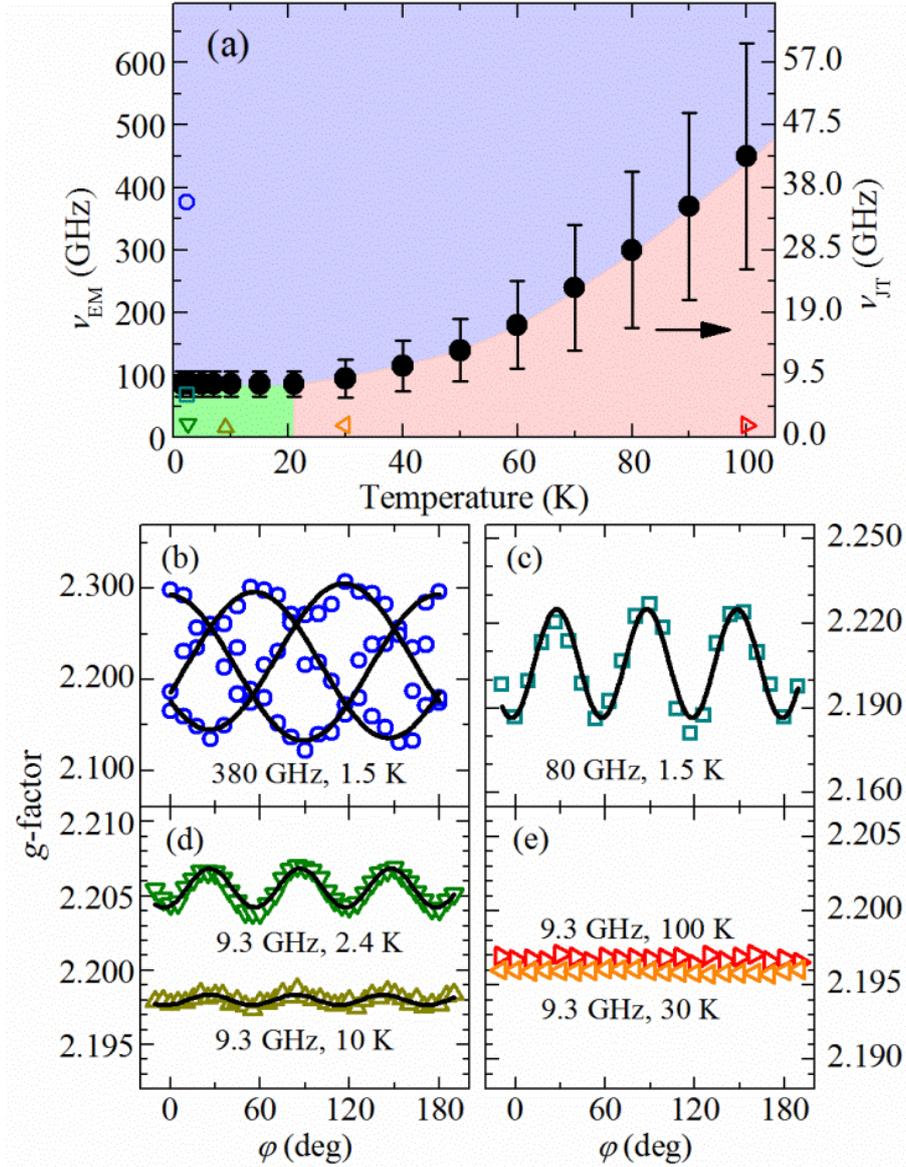

Supplemental Materials for

# Observation of the orbital quantum dynamics in the spin-1/2 hexagonal antiferromagnet Ba$_3$CuSb$_2$O$_9$

Yibo Han[1,2], Masayuki Hagiwara[1*], Takehito Nakano[3], Yasuo Nozue[3], Kenta Kimura[4], Mario Halim[4], Satoru Nakatsuji[4,5]

[1] *Center for Advanced High Magnetic Field Science, Graduate School of Science, Osaka University, 1-1 Machikaneyama, Toyonaka, Osaka 560-0043, Japan*
[2] *Wuhan National High Magnetic Field Center, Huazhong University of Science and Technology, Wuhan 430074, China*
[3] *Graduate School of Science, Osaka University, 1-1 Machikaneyama, Toyonaka, Osaka 560-0043, Japan*
[4] *Institute for Solid State Physics, University of Tokyo, 5-1-5 Kashiwanoha, Kashiwa, Chiba 277-8581, Japan*
[5] *PRESTO, Japan Science and Technology Agency (JST), 4-1-8 Honcho Kawaguchi, Saitama 332-0012, Japan*

Here we provide additional information on 1) synthesis of Ba$_3$CuSb$_2$O$_9$ single crystals, 2) details of the ESR measurements, 3) Lorentzian fitting of the X-band ESR curves, 4) evaluation of the *g*-factors for static Jahn-Teller distortion, 5) analysis of the X-band ESR intensities, 6) X-band ESR linewidth for both samples 7) ESR curves for the hexagonal sample at 80 and 380 GHz, 8) *g*-factor at different frequency and temperatures, 9) high frequency ESR curves for the orthorhombic sample 10) Schematic pictures of the orbital configuration for both samples, and 11) evaluation of the dynamic Jahn-Teller frequency.

### SM1 Synthesis of hexagonal and orthorhombic Ba$_3$CuSb$_2$O$_9$ single crystals

The Ba$_3$CuSb$_2$O$_9$ single crystals with the size of 1.2×0.8×0.3(2) mm$^3$ (hexagonal) and 0.9×0.9×0.1(2) mm$^3$ (orthorhombic) were grown by a flux method at the Institute for Solid State Physics, The University of Tokyo. A mixture of polycrystalline Ba$_3$CuSb$_2$O$_9$ and BaCl$_2$ was inserted into a Pt crucible and heated to 1500 K, followed by slow cooling. The Sb/Cu ratio was adjusted in the initial growing process. For the orthorhombic sample the Sb/Cu ratio was 2.13(1), and for the hexagonal it was 2.01(1). The crystals' quality was checked by X-ray diffraction [14]. It was found the orthorhombic sample contains an orthorhombic volume fraction of about 95%, and the hexagonal sample contains a hexagonal volume fraction larger than 99%.

### SM2 Details of the ESR measurements

For the multi-frequency ESR measurements, three types of apparatus were employed: (I) An X-band ESR apparatus (Bruker EMX ESR Spectrometer) was used for the precise measures of the temperature and angular dependent ESR spectra at $\nu_{EM}$ = 9.3 GHz. The samples were mounted on a quartz rod, which was connected with a rotator. Hence, the sample could be rotated relatively to the external magnetic field. The temperature is variable from 2.4 K to room temperature. A field derived mode was used for this apparatus. (II) A vector network analyzer (AB Millimetre Co. Ltd.) together with a 16 T superconducting magnet (Oxford Instruments Co. Ltd.) were used for the ESR measurements at the frequencies from 20 GHz to 380 GHz. For $\nu_{EM} \leq 50$ GHz, ESR cylindrical resonators were used to increase the signal-to-noise ratio (S/N), and for $\nu_{EM} > 50$ GHz, transmission-type probes were used. For most frequencies, the temperature in the sample chamber was varied from 1.5 K to about 120 K. At 380 GHz, the highest temperature was 150 K. At above 120 K/150 K, S/N becomes very small. (III) A nondestructive pulse magnet, a far-infrared laser (Edinburgh Instruments, FIRL100), and an InSb detector (QMC Instruments) were used to measure the ESR at $\nu_{EM} = 730$ GHz. The duration of the pulsed field was about 7 ms and the highest magnetic field was 55 T in the measurements. The temperature was varied from 1.3 K to 100 K. At $T > 50$ K, S/N becomes very small.

## SM3 Lorentzian fitting of the X-band ESR curves

By fitting to four field-derivative Lorentzian curves, a good accordance between the data and total fit curves was obtained for the ESR curves of the orthorhombic sample as shown in Fig. S1(a). In contrast, each spectrum of the hexagonal sample was perfectly fit by a single Lorentzian curve, as shown in Fig. S1(b). The ESR parameters including the intensity, line shape and $g$-factors are derived from these fittings.

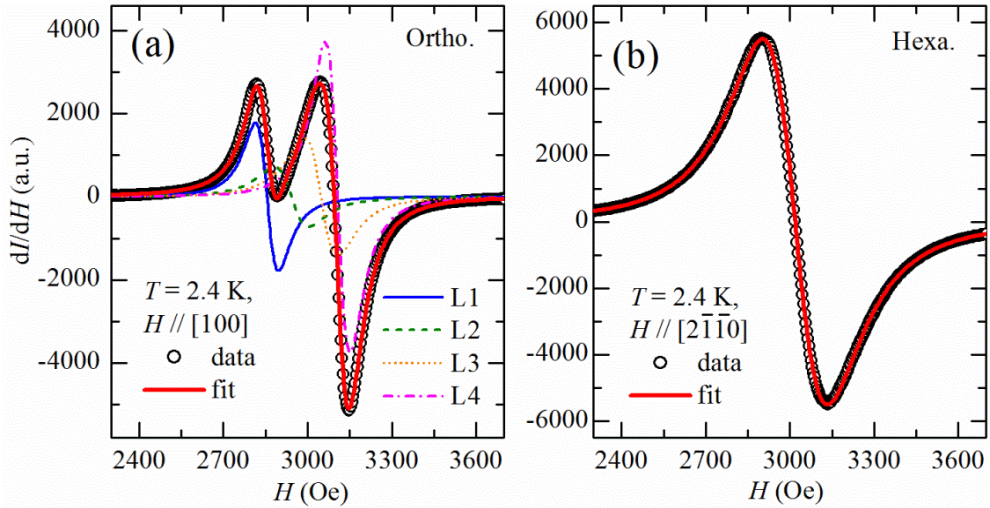

FIG. S1. Derivative Lorentzian curve fitting of the X-band ESR data. The orthorhombic is shown in (a) and the hexagonal sample in (b).

The field derivative-Lorentzian formula is: $dI/dH = 16Aw(H-H_C)/\{\pi[4(H-H_C)^2+w^2]^2\}$, where $I$ is the magnitude of ESR signal, $H$ is the external magnetic field. $A$, $w$ and $H_C$ are the intensity factor, the line width, and the resonance field, respectively. The ESR intensity is calculated by the integration $P = \int IdH$. The $g$-factor is calculated by $g = h\nu_{EM}/(\mu_B H_C)$, where $h$ is the Planck constant, and $\mu_B$ is the Bohr magneton.

## SM4 Evaluation of the $g$-factors for static Jahn-Teller distortion.

The Jahn-Teller distortion occurs in an orbital degenerate system, especially the $Cu^{2+}(d^9)$ ion in a octahedral coordination. In the case of the ground state with $3d$ ($x^2-y^2$) (hole picture), the g-factor along the elongated Cu-O direction ($g_∥$) would be larger than that at a symmetric and isotropic situation ($g_1$), and the g-factor along the compressed direction $g_⊥$ would be smaller than $g_1$. The difference between $g_∥$ and $g_1$ is defined as $g_2$. The anisotropic g-factors of the orthorhombic sample in the elongated situation of the static Jahn-Teller distortion were evaluated by [23-24],

$$g_∥ = g_1 + g_2 = 2 - 8λ/Δ'$$
$$g_⊥ = g_1 - 0.5g_2 = 2 - 2λ/Δ' \tag{S1}$$

where $λ$ is the spin-orbit coupling constant, and $Δ'$ is the energy gap between $e_g$ and $t_{2g}$ orbital states. In order to get $g_⊥$ and $g_∥$, we fit the data based on the following analysis.

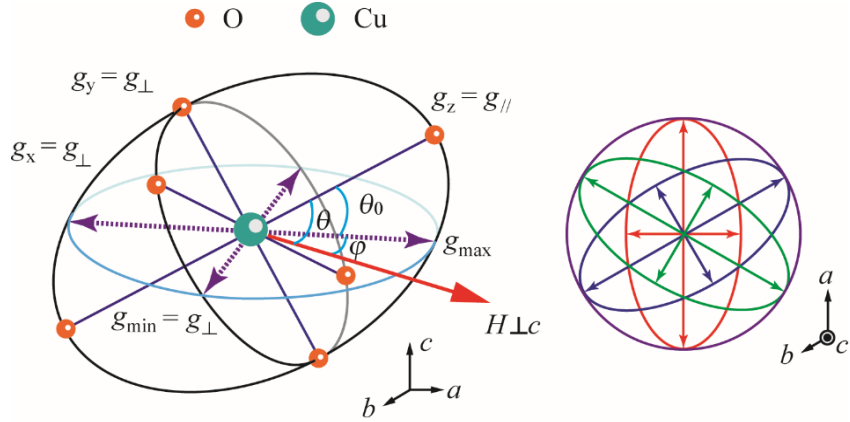

FIG. S2. g-factors for a elongated-type static Jahn-Teller distortion. The left figure shows the g-factors for an elongating $CuO_6$ octahedron, the magnetic field is perpendicular to the crystal c-axis and is rotated in the ab plane, with an angle $φ$ to a-axis. The right figure shows the g-factors for the elongations along the three respective Cu-O bond directions viewed from the c-direction.

As shown in Fig. S2, the elongated $CuO_6$ octahedron in BCSO would induce an ellipsoidal g-factor anisotropy in real space, with the characteristic g-factors $g_z=g_∥$, and $g_x = g_y = g_⊥$. When the magnetic field is rotated inside the ab plane, an elliptical g-factor distribution with two-fold anisotropy would be obtained:

$$g_{ab} = \sqrt{g_∥^2 \cos^2θ + g_⊥^2 \sin^2θ} \tag{S2}$$

where $θ$ is the angle from the local z direction to the magnetic field ($H$) direction (see Fig. S2). $θ$ is related to $θ_0$ and $φ$ as follows:

$$\sinθ = \sqrt{\sin^2θ_0 + \cos^2θ_0 \sin^2φ}$$
$$\cosθ = \cosθ_0 \cosφ \tag{S3}$$

where $θ_0$ is the angle from the local z direction to the ab plane. When we define the ratio n as n = $g_∥/g_⊥$, $g_{ab}$ obtained from equation (S2)-(S3) is:

$$g_{ab} = g_⊥ \sqrt{\frac{n^4}{n^2+1/2}\cos2φ + \frac{1}{2(n^2+1/2)} + \frac{n^2}{n^2+1/2}\sin^2φ} \tag{S4}$$

By fitting the main branches in Fig. 1(c) (main text) using equation (S4), one obtains $g_⊥ = 2.08(1)$ and $g_∥ = 2.41(1)$.

## SM5 Analysis of the X-band ESR intensities of the hexagonal sample

As the supplemental information of Fig. 2(a) in the main text, we show the detailed calculation of the X-band ESR integrated intensity ($P = \int IdH$) as a function of temperature. It was found that the ESR intensity does not show anisotropy, and thus only the data along $[2\bar{1}\bar{1}0]$ axis are shown. The $P$-$T$ relation in Fig. S3(a) is similar to the $\chi$-$T$ plot [Fig. S3(b)]. The resonance magnetic field is only 0.3 T and thus the Zeeman energy should be regarded as a small perturbation. By fitting the data to the Curie-Weiss law at low temperature region ($< 7$ K), the "dimer" component was extracted from the total intensity by subtracting the Curie-Weiss part, which indicates that part of the coppers form spin-singlet dimers below about 60 K. By taking an approximation of a simple singlet-triplet transition model, a fit to the intensity using the "gap" component at low temperatures ($\leq 60$ K) [25]

$$I = 2I_0 Z^{-1} \exp(-\Delta/k_B T)\sinh(g\mu_B H/k_B T),$$
$$Z \equiv 1+\exp(-\Delta/k_B T)\{1+2\cosh(g\mu_B H/k_B T)\}, \quad (S5)$$

yields the "singlet-triplet" gap energy $\Delta_{hexa} \sim 68$ K for the hexagonal, and $\Delta_{ortho} \sim 73$ K for the orthorhombic sample, which are similar in magnitude to the ones estimated for polycrystalline [13].

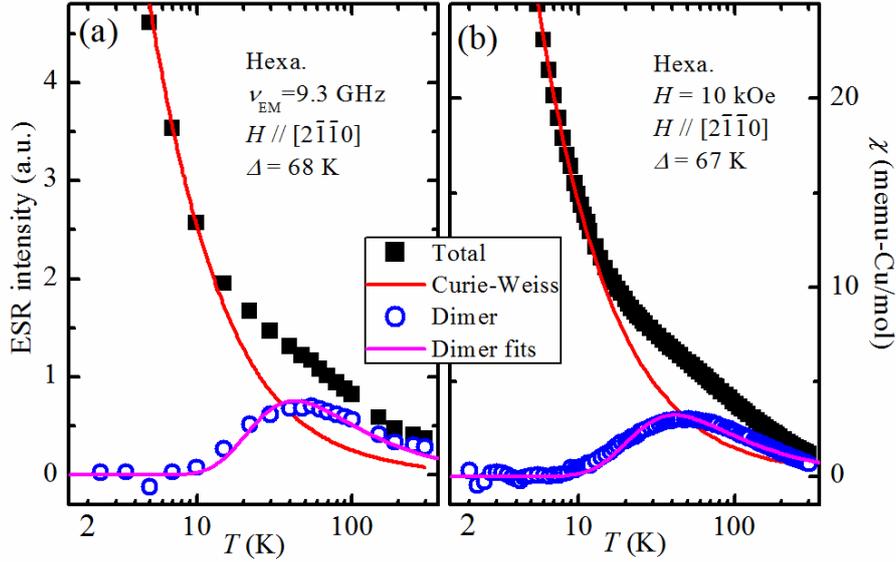

FIG S3. X-band ESR intensity and magnetic susceptibility of the hexagonal sample show similar tendency. The $\Delta$ values are estimated by fitting the dimer data using Eq. (S5).

## SM6 X-band ESR linewidth for both samples.

We show here the in-plane ($ab$ plane) X-band ESR linewidth anisotropy of both samples as the supplemental information for Fig. 2(b) in the main text. The in-plane linewidth for the hexagonal sample is almost isotropic even at the lowest temperature of 2.4 K, with a tiny six-fold-like symmetry. On the other hand, there are three sub-branches in the orthorhombic sample, the linewidths of which also show a similar periodic behavior, and are consistent with the two-fold $g$-factor symmetry shifting by 60 degrees to each other.

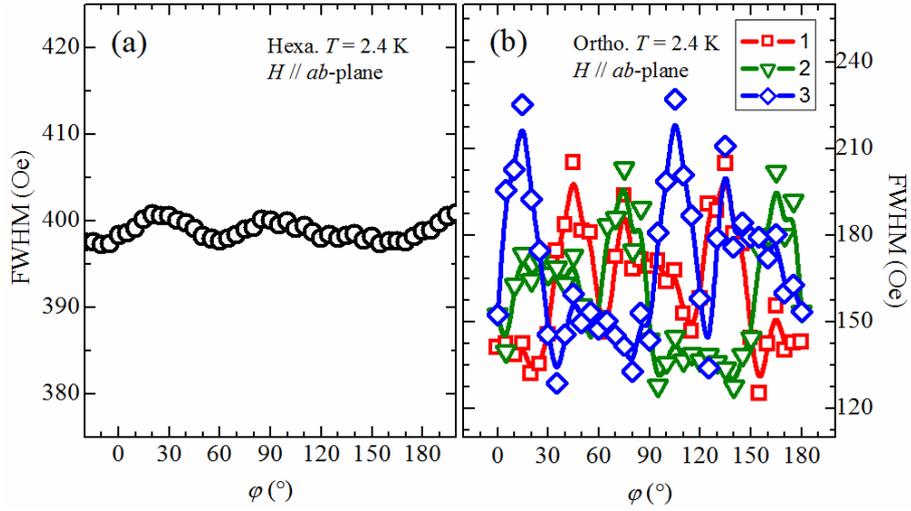

FIG. S4. X-band ESR linewidth in the *ab*-plane for (a) the single Lorentzian curve of the hexagonal sample, and (b) the three sub-Lorentzian curves of the orthorhombic.

## SM7 ESR curves at 80 and 380 GHz.

As the supplemental data for Fig. 4 in the main text, figures S5(a) and S5(b) show the angular dependent ESR curves measured at $\nu_{EM}$= 80 and 380 GHz from which the *g*-factors in Figs. 4(b), (c) in the main text were obtained.

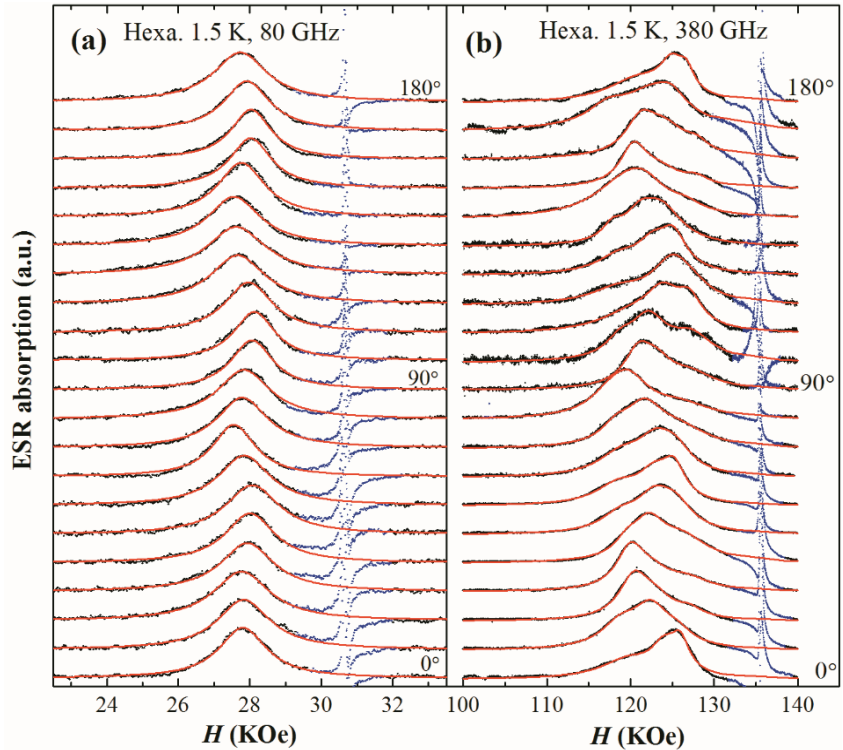

FIG. S5. ESR curves measured at (a) 80 GHz and (b) 380 GHz for the hexagonal sample. The angle $\varphi$ is defined in Fig. 1 (d) in the main text. Black dots are the experimental data and red lines are fitting curves to single (a) or triple (b) Lorentzian functions. Blue dots represent the signals of ESR standard sample DPPH, which were masked during the fitting.

**SM8 *g*-factor as a function of frequency and temperature.**

As the supplementary information for Fig. 4 in the main text, we show the complete data for the *g*-factor as a function of temperature $T$ and microwave frequency $\nu_{EM}$. One can find in Fig. S6 several sets of *g-T* plots at various frequencies $\nu_{EM}$ = 9.3, 28, 50, 80, 120, 180, and 380 GHz. The *g-T* plots for the orthorhombic sample at $\nu_{EM}$ = 9.3 GHz is placed at the rightmost side in order to clearly show its evolution with increasing $\nu_{EM}$, from the almost isotropic (red, the same as Fig. 1f in the main text) to the anisotropic patterns (black, the same as Fig. 1e in the main text), which indicates a crossover from dynamic to static Jahn-Teller distortion as discussed in the main text. Figure S4 also reveals that the six-fold symmetric *g*-factor only appears in the region of $\nu_{EM} \leq 80$ GHz and $T \leq 20$ K, where the "orbital liquid" manifests itself through the six-fold symmetric exchange pathways that dynamically switch due to the orbital tunneling.

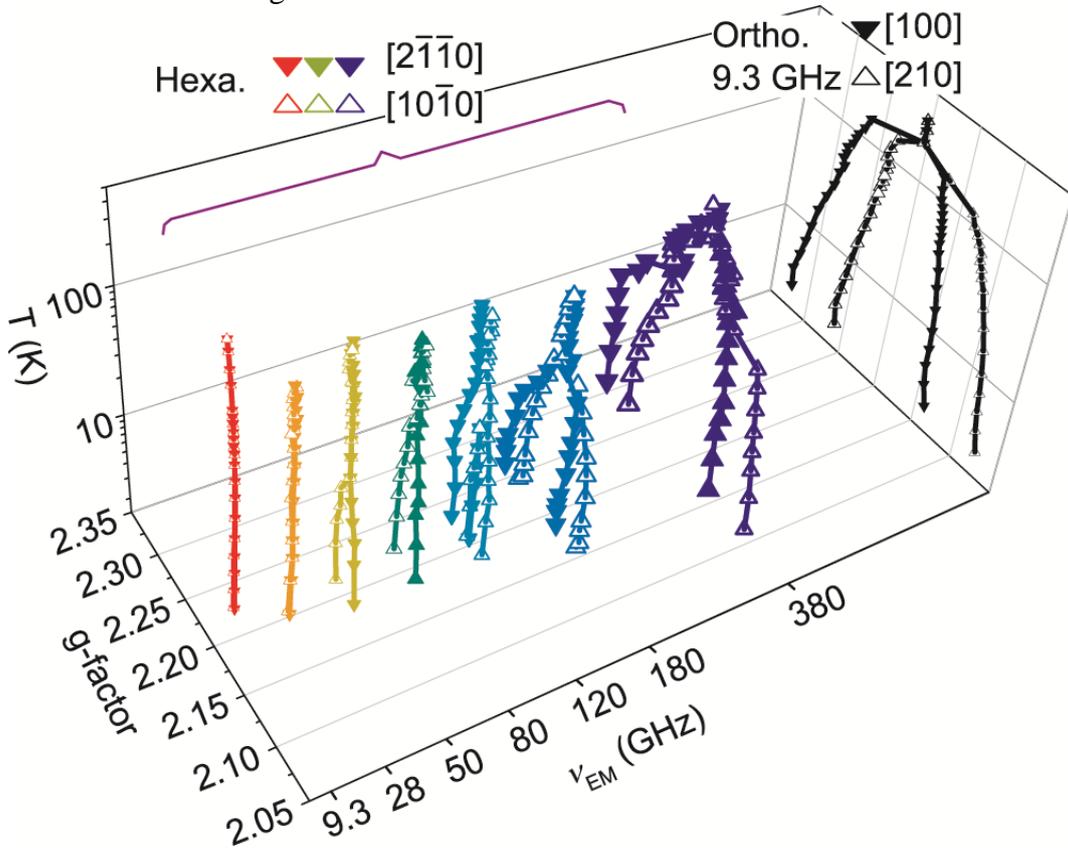

FIG. S6. *g*-factor varied with frequencies and temperature. The data from 9.3 GHz (red triangles) to 380 GHz (blue triangles) show the *g*-factors of the hexagonal sample, and the black triangles at the rightmost side shows that of the orthorhombic sample from X-band ESR results. Open and closed triangles show the *g*-factors for the external magnetic field parallel to $[2\bar{1}\bar{1}0]/[100]$ and $[10\bar{1}0]/[210]$ respectively. The solid lines connecting the data points are drawn for clarity. The error bars are shorter than the symbols' size and all the data points are obtained by Lorentzian curve fitting.

**SM9 High frequency ESR results of the orthorhombic sample.**

After showing detailed information on the *g*-factor anisotropy and lineshape of the hexagonal sample, it is also useful to demonstrate that of the orthorhombic sample, especially in the high frequency region. Figure S7 shows the frequency evolution of the ESR lineshape at the lowest temperature, which does not change so much with increasing frequency from 9.3 GHz to 300 GHz. It is obvious that the static orbital order in this sample does not induce any lineshape change when the ESR frequency turns to be higher, in sharp contrast to the hexagonal sample, for which high-frequency-induced narrowing and splitting were observed. The reason is that the orbital fluctuation is absent, and thus the ESR line shape reflects the magnetic anisotropy caused by the static JT effect.

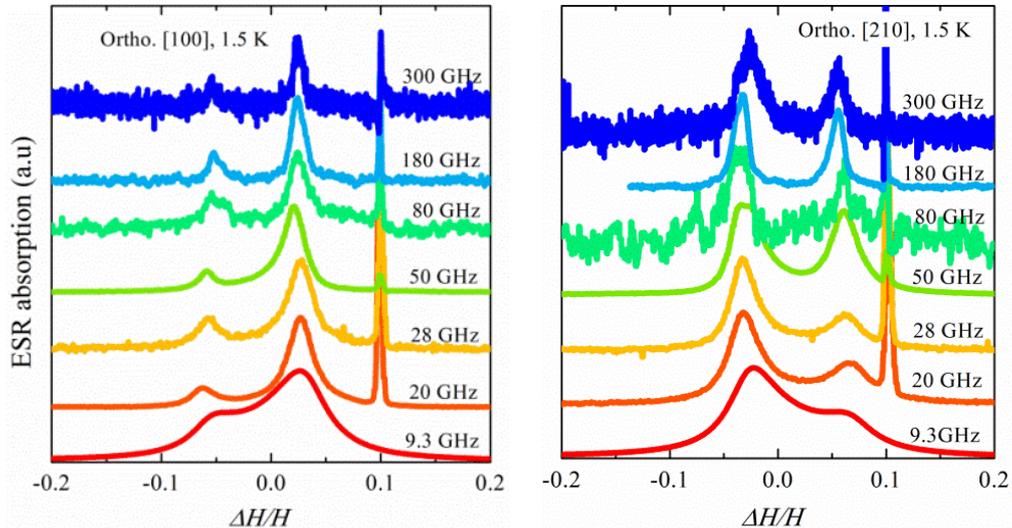

FIG. S7. Frequency dependent ESR curves for the orthorhombic sample at $T = 1.5$ K along [100] and [210] directions. The sharp peaks at $\Delta H/H = 0.1$ is the signal from DPPH.

**SM10 Orbital configurations in the orthorhombic and hexagonal samples.**

To clearly show the conclusion derived from our ESR results, schematic pictures of orbital configurations are shown in the figures below.

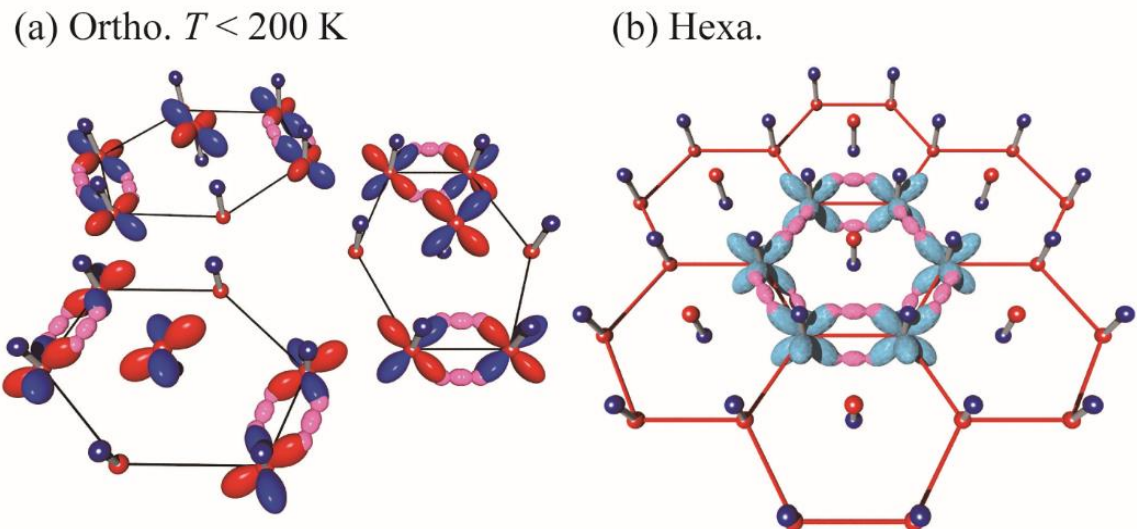

FIG. S8. Schematic pictures of the orbital configuration for both samples conjectured based on the X-band ESR results at the lowest temperatures. The Cu-Sb dumbbells form honeycomb-like lattices with Cu atoms as red balls and Sb as blue ones. (a) The ferro-orbital ordered states in the orthorhombic sample. The [Cu] $3d$ ($x^2-y^2$) ($3d$ ($y^2-z^2$), $3d$ ($z^2-x^2$)) orbitals are shown by blue and red ellipsoids and [O] $2p(z)$ orbitals by pink ones. (b) The possible dynamic orbital states in the hexagonal sample, which form resonating singlet dimers in the honeycomb-based lattice. The light blue ovals show the temporal overlapping of the [Cu] $3d$ ($x^2-y^2$), $3d$ ($y^2-z^2$), and $3d$ ($z^2-x^2$) orbitals.

## SM11 Evaluation of the dynamic Jahn-Teller frequency.

Here we show the calculation of the dynamic Jahn-Teller frequency $\nu_{JT}$ for the hexagonal sample, with the data obtained from both samples.

The anisotropic Zeeman energy is defined as $E_Z' = g_2\mu_B H$, in which $g_2$ is evaluated from equation (S1), one can get $\lambda/\Delta' = -0.05$, $g_1 = 2-4\lambda/\Delta' = 2.20$, $g_2 = -4\lambda/\Delta' = 0.21$, and $E_Z' = g_2\mu_B H$.

The Jahn-Teller frequency $\nu_{JT} = 1/\tau \approx g_2\mu_B H_C/h$ is obtained, because when $\tau \ll h/(g_2\mu_B H_C)$, a motional narrowing takes place, and when $\tau \gg h/(g_2\mu_B H_C)$, an anisotropic g-factor corresponding to static Jahn-Teller distortion should be observed [23-24]. Hence, $\nu_{JT} = 1/\tau \approx g_2\mu_B H_C/h = 0.21*\mu_B H_C/h$. From $H_C = h\nu_C/(g_1\mu_B) = h\nu_C/(2.2*\mu_B)$, where $\nu_C$ is the critical microwave frequency above which the splitting occurs, $\nu_{JT} = \nu_C g_2/g_1 \approx 0.095*\nu_C$ is extracted.